# Observation of Fermi-energy dependent unitary impurity resonances in a strong topological insulator $Bi_2Se_3$ with scanning tunneling spectroscopy

M. L. Teague[1], H. Chu[1], F.-X. Xiu[2,3], L. He[2], K.-L. Wang[2], N.-C. Yeh[1,4*]

[1]*Department of Physics, California Institute of Technology, Pasadena, CA 91125, USA*

[2]*Department of Electrical Engineering, University of California, Los Angeles, CA 90095, US*

[3]*Department of Electrical and Computer Engineering, Iowa State University, Ames, IA 50011, USA*

[4]*Kavli Nanoscience Institute, California Institute of Technology, Pasadena, CA 91125, USA*

**Abstract**

Scanning tunneling spectroscopic studies of $Bi_2Se_3$ epitaxial films on Si (111) substrates reveal highly localized unitary impurity resonances associated with non-magnetic quantum impurities. The strength of the resonances depends on the energy difference between the Fermi level ($E_F$) and the Dirac point ($E_D$) and diverges as $E_F$ approaches $E_D$. The Dirac-cone surface state of the host recovers within $\sim$ 2Å spatial distance from impurities, suggesting robust topological protection of the surface state of topological insulators against high-density impurities that preserve time reversal symmetry.

An exciting development in modern condensed matter physics is the beautiful manifestation of topological field theories in strongly correlated electronic systems, where topological field theories [1] are shown to provide a classification of order due to macroscopic entanglement that is independent of symmetry breaking [2]. The fractional quantum Hall (FQH) state is the first known example of such a quantum state that exhibits no spontaneous broken symmetry and has properties depending only on its topology rather than geometry [2]. Recently, a new class of time-reversal symmetry protected topological states known as the quantum spin Hall (QSH) states or the topological insulators (TI) has emerged and stimulated intense research activities [3,4].

One of the novel properties associated with the TI is the presence of a Dirac spectrum of chiral low-energy excitations, which is a salient feature of the Dirac materials that exploits the mapping of electronic band structures and an embedded spin or pseudo-spin degree of freedom onto the relativistic Dirac equation [3–9]. These materials, including graphene [9] and the surface state (SS) of three dimensional (3D) strong topological insulators (STI) [3–8], have emerged as a new paradigm in condensed matter for investigating the topological phases of massless and massive Dirac fermions. In the case of 3D-STI, an odd number of massless Dirac cones in their SS is ensured by the $Z_2$ topological invariant of the fully gapped bulk [3–8]. Backscattering of Dirac fermions is suppressed due to topological protection that preserves the Dirac dispersion relation for any time-reversal invariant perturbation [3,4]. Thus, 3D-STI are promising materials for applications in areas of spintronics [3,4,10] and topological quantum computation [3,4,11] if their SS exhibit sufficient stability to impurities [12,13].

While direct backscattering is prohibited in both the SS of 3D-STI and in graphene, sharp resonances are not excluded because Dirac fermions with a finite parallel momentum may be confined by potential barriers [9]. In fact, theoretical calculations for Dirac fermions in the presence of non-interacting impurities have predicted the occurrence of strong impurity resonances [12, 13]. Nonetheless, no direct empirical observation of strong resonances has been demonstrated to date despite numerous reports of spectral evidences for quasiparticle interferences associated with impurity or step-edge induced scattering [14–16]. In this letter we report direct scanning tunneling spectroscopic (STS) observation of impurity resonances in a 3D-STI system, $Bi_2Se_3$. We find that the strength of non-magnetic impurity resonances appears strongly dependent on the energy difference between the Fermi level ($E_F$) and the Dirac point ($E_D$) and diverges as $E_F \to E_D$. The impurity resonances occur near $E_D$ and are localized within a small region of a radius $r_0 \sim$ 2Å so that the SS spectra of the host remain undisturbed even for high-density unitary impurities. These findings suggest that the SS of a 3D-STI is topologically well protected against impurities that preserve time reversal symmetry. Moreover, the absence of strong impurity resonances in other spectroscopic studies [14–17] may be attributed to the large energy difference between $E_F$ and $E_D$ that led to substantial screening of the impurity states.

---

[*] Corresponding author. E-mail address: ncyeh@caltech.edu

The samples investigated in this work are epitaxial $Bi_2Se_3$ films grown on Si(111) by molecular beam epitaxy (MBE). Details of the growth process have been described elsewhere [18]. Transmission electron microscopy (TEM) on these films exhibited perfect triangular lattice structures, and ARPES (angle resolved photoemission spectroscopy) studies revealed a single Dirac cone [18]. Figure 1(a) shows an atomic force microscope (AFM) image of an as-grown $Bi_2Se_3$ epitaxial film with an average thickness of 44 quintuple layers (QLs). The film surface consists of large triangle-shaped flat terraces, reflecting the hexagonal crystalline structure inside the (0001) plane. The height of each terrace is ~ 0.95 nm, corresponding to a single QL thickness. The typical lateral dimension of the top layer ranges from 150 to 350 nm, and the width of each subsequent terrace is 70 ~ 90 nm.

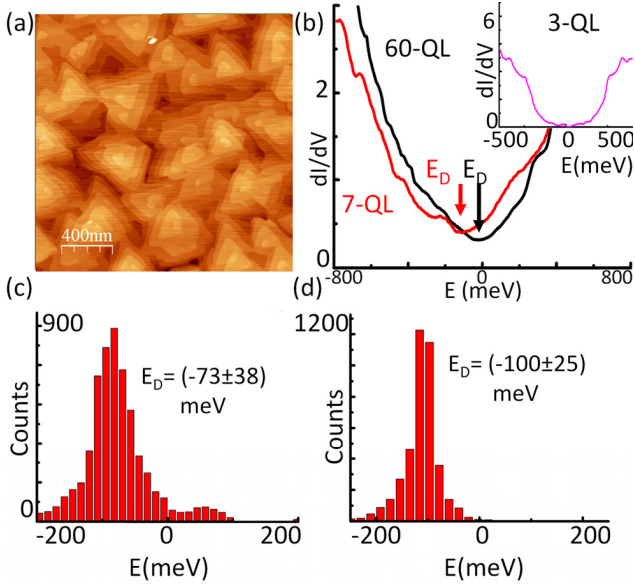

Fig. 1. Characteristics of MBE-grown $Bi_2Se_3$ epitaxial films on Si(111): **(a)** AFM image of a sample with a mean thickness of 44 QLs, showing triangle-shaped flat terraces. **(b)** Main panel: Comparison of the typical tunneling conductance spectra of two $Bi_2Se_3$ films of 60-QL and 7-QL thicknesses. The Dirac point $E_D$ shifts away from the Fermi level $E_F = 0$ with decreasing thicknesses. Inset: A representative tunneling spectrum for a 3-QL sample, showing opening of an energy gap around $E_F$. **(c)** Histogram of $E_D$ in the 60-QL sample. **(d)** Histogram of $E_D$ in the 7-QL sample comparable results.

After MBE-growth, samples were transferred to the cryogenic probe of a homemade scanning tunneling microscope (STM). The sealed STM assembly was evacuated and cooled to either 6 K or 77 K in ultra-high vacuum. Both spatially resolved topography and normalized tunneling conductance $(dI/dV)/(I/V)$ vs. energy ($E = eV$) spectroscopy were acquired pixel-by-pixel simultaneously, with tunneling currents perpendicular to the sample surface, and the typical junction resistance was ~ 1 GΩ. Detailed survey of the surface topography and tunneling conductance spectra was carried out over typically (8×8) nm$^2$ areas, and each area was subdivided into (128×128) pixels.

Generally the normalized tunneling conductance spectra in our STS studies were found consistent throughout a flat area. Representative point spectra for the 60-QL and 7-QL samples are given in the main panel of Fig. 1(b), and the ranges of the Dirac energy for all areas investigated are $E_D$ = (−73±38) meV and $E_D$ = (−100±25) meV, respectively, as summarized by the histograms of the Dirac energies shown in Figs. 1(c) and 1(d). In contrast, a typical spectrum for the 3-QL sample (inset of Fig. 1(b)) reveals apparent opening of an energy gap (~ 0.4 eV) around $E_F = 0$.

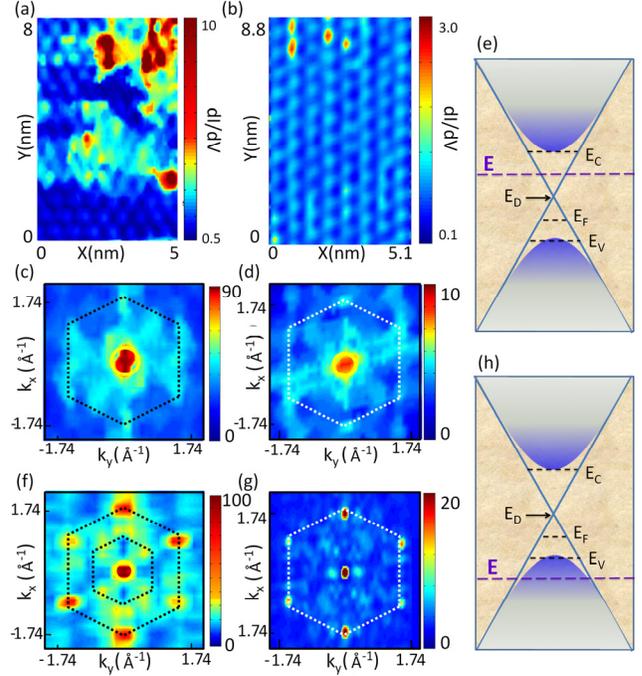

Fig. 2. (Color online) **(a)** Atomically resolved ($dI/dV$) map of a 60-QL sample at $E = −89$ meV. **(b)** Atomically resolved ($dI/dV$) map of a 7-QL sample at $E = −130$ meV. **(c)** Fourier transformation (FT) of the ($dI/dV$) map for $E = −30$ meV over the same area shown in (a) for the 60-QL sample, showing a circular diffraction ring consistent with the SS dispersion relation for $E_v < E < E_c$ where $E_v$ and $E_c$ refer to the top of the bulk valence band and the bottom of the bulk conduction band, respectively, and ($E_c−E_v$) ~ 300 meV, and the reciprocal space units are given in the convention of (2π) over lattice constants. **(d)** FT conductance map for $E = −50$ meV over the same area shown in (b) for the 7-QL sample, showing a circular diffraction ring consistent with the SS dispersion relation for $E_v < E < E_c$. **(e)** Schematic illustration of the energy dispersion relations associated with the bulk and surface states of the 3D-STI $Bi_2Se_3$, showing an apparently circular Fermi surface for $E_v < E < E_c$. **(f)** FT conductance map for $E = −300$ meV over the same area shown in (a) for the 60-QL sample, showing clear first-order and weak second-order Bragg diffraction spots for $E < E_v$. **(g)** FT conductance map for $E = −300$ meV over the same area shown in (b) for the 7-QL sample, showing first-order Bragg diffraction spots for $E < E_v$. **(h)** Schematic energy dispersion relations associated with the bulk and surface states of the 3D-STI $Bi_2Se_3$, showing dominating bulk contributions for $E < E_v$.

Despite the relatively consistent tunneling spectra for most areas in view, we note the presence of a few atomic impurities, as manifested by the localized high conductance spots in Fig. 2(a) for the 60-QL sample and in Fig. 2(b) for the 7-QL sample. On the other hand, unlike other 3D-STI (e.g., $Bi_2Te_3$ and $Bi_{1-x}Sb_x$) with more complicated Fermi surfaces that lead to SS deviating from a perfect Dirac cone as well as impurity-induced quasiparticle interferences (QPI) for sufficiently high-energy quasiparticles [14–16], the SS of $Bi_2Se_3$ does not exhibit discernible QPI because the perfect single Dirac cone prevents backscattering. Hence, the Fourier transformation (FT) of the conductance maps primarily exhibited round spots in the center of the FT conductance maps and very faint Bragg diffraction spots at low energies within the surface state, as shown in Figs. 2(c)-(e). Eventually significant Bragg diffraction peaks appear in the reciprocal space for energies merged into the bulk state, as manifested in Figs. 2(f)-(h) for the FT conductance maps taken at $E = -300$ meV on the 60-QL and 7-QL samples, respectively. In contrast, QPI with wave-vectors smaller than the reciprocal lattice constants were found in samples thinner than 6 QLs [21] due to modified SS as the result of wave-function overlapping and Rashba-type spin-orbit splitting between the top and bottom surfaces of the thin film [19,20]. Given that the energy dispersion relation for the 3-QL sample deviates from that of a Dirac cone, in the following we focus our studies of the impurity resonances only on the 60-QL and 7-QL samples.

To investigate the spectral evolution associated with the presence of these quantum impurities, we show in Figs. 3(a)-(f) different line-cuts across a (5×8) nm$^2$ constant-bias conductance map and the corresponding spectra for the 60-QL sample. For a line-cut along an area without impurities as exemplified in Fig. 3(a), the tunneling spectra are generally consistent everywhere, showing a Dirac energy $E_D = (-35\pm10)$ meV slightly below the Fermi level. On the other hand, the tunneling spectra directly above quantum impurities reveal strong resonant conductance peaks at $E \sim E_D$. Moreover, these resonant peaks are spatially confined to a region of $\sim 2$Å in radius, as shown by the spectra along various line-cuts in Figs. 3(b)-(d) and further elaborated in Fig. 5(a). These spectral characteristics clearly reveal that the SS of the host recovers rapidly from impurities.

Interestingly, for quantum impurities separated by only one lattice constant, the spectral characteristics for the inter-impurity region exhibit strong interferences for energies deep into the bulk valence band while the SS spectra have restored to that of the host, as exemplified in Figs. 3(e)-(f). These findings therefore imply strong topological protection of the SS against impurities even in the limit of significant effects on the bulk state. For comparison, while similar quantum impurities are observed in the 7-QL sample, the intensity of the impurity resonances is much reduced, as illustrated in Figs. 4(a) and 4(b). As discussed later, this weakened impurity resonance may be attributed to the larger energy separation $(E_F - E_D) \equiv \tilde{E}$ (see Fig. 1(b) and Fig. 4(c)).

To better quantify the spatial confinement and energy dependence of the impurity resonance, we illustrate in Fig. 5(a) the spatial dependence (r) of the tunneling conductance near one of the isolated impurities in the 60-QL sample. For $E \sim E_D$, we find strong resonance in the tunneling conductance over a very narrow spatial range $r \sim \pm 2$Å, as illustrated by the solid curve in Fig. 5(a). On the other hand, for $E < E_D$ but still within the SS, the spectral resonance diminishes rapidly, as shown by the curve of black symbols in Fig. 5(a). Similarly, no impurity resonance is visible for energies deep into the bulk valence band, as exemplified by the $(dI/dV)$-vs.-$r$ curve taken at $E = -275$ meV (red symbols) in Fig. 5(a). In the case of two closely located impurities, we find that the impurity resonances at $E \sim E_D$ remains strongly localized spatially (solid blue curve in Fig. 5(b)). Moreover, the SS spectrum of the small intermediate region between two impurities appears to fully recover to that of the host (black symbols in Fig. 5(b)), whereas the bulk spectrum ($E = -400$ meV) for the same intermediate region exhibits strong interferences, as exemplified by the red symbols in Fig. 5(b) and also in Fig. 3(e). The rapid recovery of the SS spectrum from impurities may be understood as the result of topological protection of the SS in $Bi_2Se_3$, even in the limit of high-density impurities.

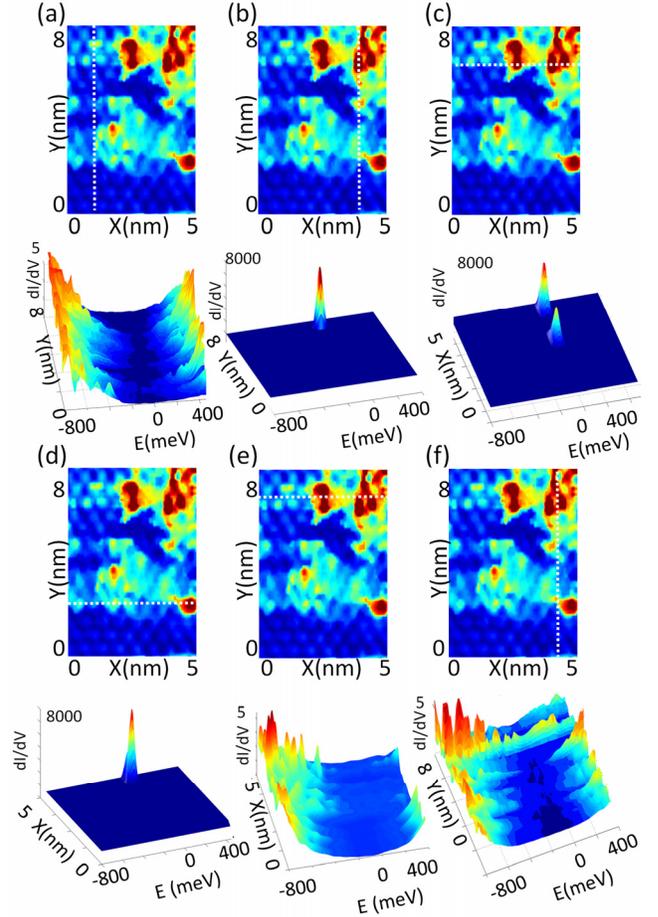

Fig. 3. (Color online) Spectral evolution along various line-cuts of a (5×8) nm$^2$ area of a 60-QL sample at $T = 77$ K, where the white dotted line in each upper panel represents a line-cut across an atomically resolved constant-energy conductance $(dI/dV)$ map, and the corresponding $(dI/dV)$-vs.-$E$ spectra along the line-cut are given in the lower panel: **(a)** Across an impurity-free region; **(b)** Across a single impurity; **(c)** Across two impurities; **(d)** Across an isolated impurity; **(e)** Between two closely spaced impurities along the horizontal direction; **(f)** Between two closely spaced impurities along the vertical direction.

Similarly, for the 7-QL sample with a larger value of $|\tilde{E}|$, the impurity resonance at $E \sim E_D$ for either an isolated impurity or two closely spaced impurities is also highly localized, as exemplified in Figs. 5(c)-(d). Moreover, the SS spectrum recovers rapidly and the effect of adjacent impurities on the bulk valence band diminishes significantly (Fig. 5(d)) relative to that of the 60-QL sample (Fig. 5(b)), probably due to stronger screening associated with a larger $|\tilde{E}|$ value in the 7-QL sample.

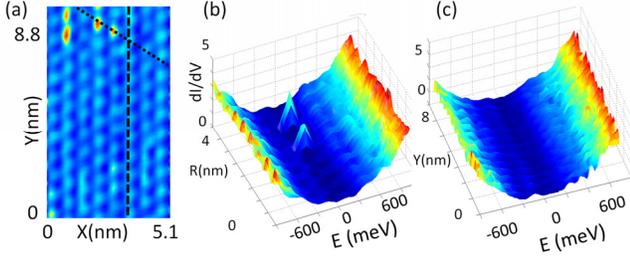

Fig. 4. (Color online) Spectral evolution along various line-cuts of a (5.1×8.8) nm$^2$ area of a 7-QL sample at $T$ = 77 K: **(a)** Atomically resolved constant-bias conductance map for $E$ = 5 meV; **(b)** ($dI/dV$)-vs.-$E$ spectra along the slanted dotted line in (a) that cuts across two point impurities; **(c)** ($dI/dV$)-vs.-$E$ spectra along an impurity-free region represented by the vertical dashed line in (a).

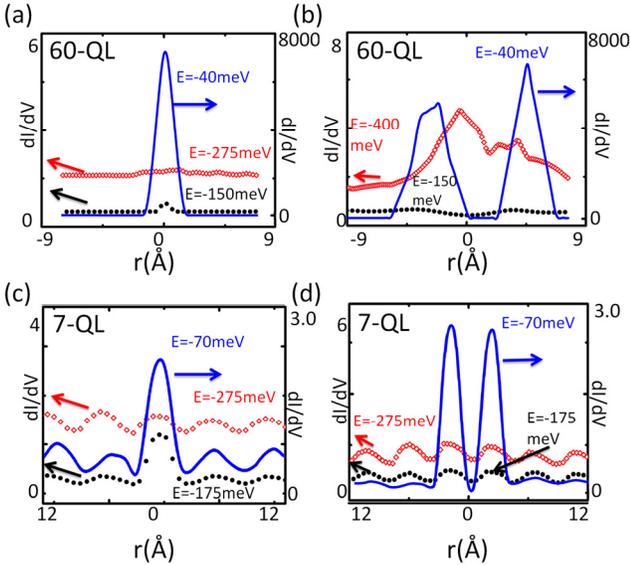

Fig. 5. (Color online) Spatial distribution and energy dependence of the impurity resonances for 60-QL and 7-QL samples: **(a)** ($dI/dV$) vs. spatial distance ($r$) spectrum of a 60-QL sample from the center of an isolated impurity for $E \sim E_D$ (blue solid curve), $< E_D$ (within the SS, black solid circles) and $\ll E_D$ (in the bulk valence band, red open diamonds). **(b)** ($dI/dV$)-vs.-$r$ spectrum of a 60-QL sample from the center of two adjacent impurities for $E \sim E_D$ (blue solid curve), $< E_D$ (black solid circles) and $\ll E_D$ (red open diamonds). **(c)** ($dI/dV$)-vs.-$r$ spectrum of a 7-QL sample from the center of an isolated impurity for $E \sim E_D$ (blue solid curve), $< E_D$ (black solid circles) and $\ll E_D$ (red open diamonds). All spectra reveal slight conductance modulations associated with the underlying atomic lattice structure. **(d)** ($dI/dV$)-vs.-$r$ spectrum of the 7-QL sample from the center of two adjacent impurities for $E \sim E_D$ (blue solid curve), $< E_D$ (black solid circles) and $E \ll E_D$ (red open diamonds).

To understand the quantitative dependence of impurity resonances on $\tilde{E}$, we follow the Keldysh Green function formalism detailed in Ref. [13] for tunneling conductance above a non-magnetic impurity in graphene, which may be applied to the SS tunneling conductance $g_{imp}$ in 3D-STI by reducing the conductance contributions from two sublattices in graphene to one Dirac cone in Bi$_2$Se$_3$. Specifically, the Hamiltonian for the low energy Dirac quasiparticles of topological insulators may be modeled by considering the following contributions [13]:

$$\mathcal{H} = \mathcal{H}_{TI} + \mathcal{H}_{imp} + \mathcal{H}_{tip} + \mathcal{H}_{TI\text{-}imp} + \mathcal{H}_{tip\text{-}TI} + \mathcal{H}_{tip\text{-}imp}, \quad (1)$$

where $\mathcal{H}_{TI} = (\boldsymbol{\sigma} \cdot \boldsymbol{p})$ is the Dirac Hamiltonian for the SS of a 3D-STI (with $\boldsymbol{\sigma}$ and $\boldsymbol{p}$ denoting the spin and momentum operators, respectively), $\mathcal{H}_{imp}$ is the impurity Hamiltonian, $\mathcal{H}_{tip}$ is the Hamiltonian for the STM tip, and the Hamiltonians $\mathcal{H}_{TI\text{-}imp}$, $\mathcal{H}_{tip\text{-}TI}$ and $\mathcal{H}_{tip\text{-}imp}$ describe hopping between TI and the impurity electrons, between TI and the STM tip electrons, and the STM tip electrons and the impurity, respectively [13]. Given the Hamiltonian $\mathcal{H}$ in Eq. (1), the time ($t$) dependent tunneling current $\mathcal{I}(t)$ may be expressed by the formula:

$$\mathcal{I}(t) = e\langle dN_{tip}/dt \rangle = ie\langle [\mathcal{H}, N_{tip}] \rangle/\hbar, \quad (2)$$

where $N_{tip}$ denotes the number operator of the tip electrons. Assuming non-interacting Dirac fermions and non-interacting impurities, and taking a cutoff energy $\Lambda$ beyond which the bulk states dominate, the $g_{imp}$ vs. $\omega \equiv (E/\Lambda)$ spectrum may be derived from Eq. (1) and Eq. (2) and by using the Keldysh Green functions [13]. Thus, at $T = 0$ the tunneling conductance $g_{imp}(\omega)$ above a nonmagnetic impurity in the single Dirac cone system Bi$_2$Si$_3$ becomes [13]:

$$g_{imp} = \frac{d\mathcal{I}}{dV},$$

$$= \frac{2e^2 \rho_{tip}}{h} \frac{|B(E)|^2}{\text{Im}[\Sigma_{imp}(E)]} \frac{|q(E)|^2 - 1 + 2\text{Re}[q(E)]\chi(E)}{[1+\chi^2(E)]}, \quad (3)$$

where $\rho_{tip}$ is the density of states of the STM tip, and $\Sigma_{imp}(E)$ is the self-energy of impurity. The quantities $B(E)$, $q(E)$ and $\chi(E)$ in Eq. (3) are related to the unperturbed retarded Green function of Dirac fermions $\mathcal{G}_\sigma^{(0),R}$ via the following relations [13]:

$$q \equiv \left[(W^0/U^0) + V^0 I_1(E)\right]/\left[V^0 I_2(E)\right],$$

$$\chi \equiv \left[E - E_{imp} - \text{Re}\{\Sigma_{imp}(E)\}\right]/\left[\text{Im}\{\Sigma_{imp}(E)\}\right],$$

$$B(E) \equiv U^0 V^0 I_2(E),$$

$$I_1(E) \equiv \sum_{\mathbf{k}} \text{Tr}\left\{\text{Re}\left[\mathcal{G}_\sigma^{(0),R}(E,\mathbf{k})\right]\right\},$$

$$I_2(E) \equiv \sum_{\mathbf{k}} \text{Tr}\left\{\text{Im}\left[\mathcal{G}_\sigma^{(0),R}(E,\mathbf{k})\right]\right\}, \quad (4)$$

where $\sigma$ denotes the spin index [13]. In Eqs. (3) and (4) the parameters $U^0$, $V^0$ and $W^0$ correspond to the interaction energies between the STM tip and the host TI, between the impurity and the TI, and between the STM tip and the impurity, respectively [13]. Introducing the dimensionless parameters $\delta \equiv \omega + (\varepsilon_F - \varepsilon_D)$, $\varepsilon_F \equiv (E_F/\Lambda)$, $\varepsilon_D \equiv (E_D/\Lambda)$, $u^0 \equiv (U^0/\Lambda)$, $v^0 \equiv (V^0/\Lambda)$, $w^0 \equiv (W^0/\Lambda)$, and $\omega_{imp} \equiv (\Omega_{imp} - E_D)/\Lambda$, and imposing the condition $|\delta| < 1$ so that the energy range of impurity resonance spectra is restricted to that of the SS, we simplify Eqs. (3) and (4) into the following expressions:

$$g_{imp}(\omega) \propto |\delta|(u^0)^2 \left[ \frac{|q|^2 - 1 + 2q\chi}{1 + \chi^2} \right], \quad (5)$$

where $q$ and $\chi$ are given by

$$q \equiv \left[ (w^0/u^0) + 4|\delta|v^0 \left( 2\ln|\delta| - \ln|1 - \delta^2| \right) \right] / \left( 4\pi v^0 |\delta| \right),$$

$$\chi \equiv \left[ \frac{\omega - \omega_{imp}}{4\pi |\delta| v_0^2} \right] - \left( 2\ln|\delta| - \ln|1 - \delta^2| \right). \quad (6)$$

In the limit of $|\delta| \to 0$ and for unitary impurities where $\omega_{imp} \to 0$, $g_{imp}$ diverges with an asymptotic form $[|\delta|(\ln|\delta|)^2]^{-1}$.

Using Eqs. (5) and (6), we illustrate the impurity resonant spectra for ($\tilde{E}/\Lambda$) = 0, 0.1 and 0.3 in Fig. 6(a), where we have taken $\omega_{imp} = 0$ and $T = 0$. For comparison, we illustrate in Fig. 6(b) two empirical impurity resonant spectra taken from the 60-QL and 7-QL samples together with their respective theoretical simulations in Fig. 6(c), where thermal smearing at $T = 77$ K has been included in the theoretical curves. We find that the theoretical peak positions are consistent with $\omega_{imp} \to 0$ (with $\omega_{imp} = 0$ for the 60-QL sample and $\omega_{imp} = 0.002$ for the 7-QL sample) so that the impurity resonant energy $\Omega_{imp}$ for both samples nearly coincides with the corresponding Dirac energy $E_D$. This finding suggests that the impurity resonances for both samples are in the unitary limit [12], where the impurity strength $U_{imp}$ for $\Omega_{imp} \to E_D$ diverges via the relation $(\Omega_{imp} - E_D) \sim 5\, \text{sgn}(U_{imp})/(|U_{imp}| \ln|U_{imp}|)$. We further note that empirically the impurity resonant peak positions for both samples also nearly coincide with their respective Dirac energies obtained from regions without impurities.

While qualitative and semi-quantitative understanding can be achieved with the analysis outlined above, we find that the linewidths of the experimental data are generally broader than those of the theoretical curves, and the ratio of the relative peak heights also differs between theory and experiments. These quantitative discrepancies suggest that the simple non-interacting Dirac fermion model may not completely account for our experimental findings.

More specifically, the aforementioned theoretical analysis of our experimental spectra has the following physical implications. First, the strong dependence of impurity resonances on ($\tilde{E}/\Lambda$) is a direct consequence of the linear dispersion relation of the surface Dirac fermions, which gives rise to an approximate logarithmic divergence in the limit of $E_F \to E_D$ [12,13]. In contrast, for samples with large $|\tilde{E}|$ due to excess doping, the spectral weight of impurity resonances may become too small to resolve directly with STS studies [17]. Second, the occurrence of strong resonance peaks at $E_D$ implies that these non-magnetic impurities are in the unitary limit [12]. Finally, in the $E_F \to E_D$ limit the broader linewidth and higher intensity of the experimental resonance peak than theoretical predictions [12,13] may imply the necessity to consider interacting Dirac fermions when the fermion density of states approaches zero.

In summary, we have demonstrated scanning tunneling spectroscopic evidence of impurity resonances in the surface state of a strong topological insulator, $Bi_2Se_3$. The impurities are in the unitary limit and the spectral resonances are localized spatially (within a radius ∼ 2Å). The spectral weight of impurity resonances diverges as the Fermi energy approaches the Dirac point, and the rapid recovery of the surface state from non-magnetic impurities suggests robust topological protection against perturbations that preserve time-reversal symmetry.

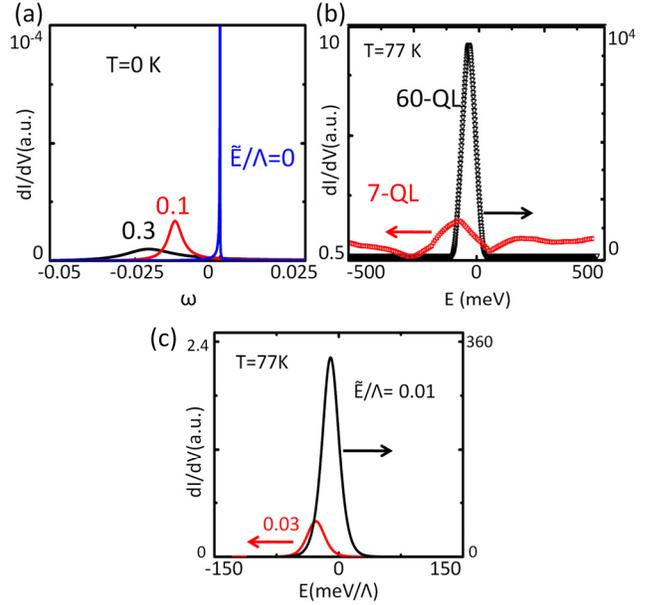

Fig. 6. Dependence of non-magnetic impurity resonances on $\tilde{E}$ in a Dirac material: **(a)** Simulated STS on top of a non-magnetic impurity for ($\tilde{E}/\Lambda$) = 0, 0.1, 0.3 and $T = 0$. The parameters used for calculations are similar to those in Ref. [13]: $\omega_{imp} = 0$, $u^0 = 0.00025$, $v^0 = 0.05$ and $w^0 = 0.0005$. **(b)** Comparison of representative empirical impurity resonance spectra of the 60-QL and 7-QL samples. **(c)** Theoretical curves generated by using Eq. (3) and the parameters $\omega_{imp} = 0$ (−0.002) and ($\tilde{E}/\Lambda$) = 0.01 (0.03) for the 60-QL (7-QL) samples. We have taken $T = 77$ K and $\Lambda = 3.0$ eV for creating the spectra in (c), and have used the same values of $u^0$, $v^0$ and $w^0$ as those in (a). We further note that the absolute values of the tunneling conductance are shown in arbitrary units so that only the relative values of the tunneling conductance under varying conditions are physically significant.

**Acknowledgements**

This work at Caltech was jointly supported by the Center on Functional Engineered Nano Architectonics (FENA), the Institute for Quantum Information and Matter, an NSF Physics Frontiers Center with support of the Gordon and Betty Moore Foundation, and the Kavli Foundation through the Kavli Nanoscience Institute (KNI) at Caltech. The work at UCLA was supported by FENA. We thank K. Sengupta and A. V. Balatsky for valuable discussion.